\documentclass[aps,prb,preprint,superscriptaddress,12pt]{revtex4-2}

\usepackage{graphicx} 
\usepackage{textcomp} 

\begin{document}
\preprint{Preprint}

\title{Infrared-induced ultrafast melting of nanostructured platinum films probed by an X-ray free-electron laser}

\author{Luca Gelisio}
\email[corresponding author:\,]{luca.gelisio@desy.de}
\affiliation{Deutsche Elektronen-Synchrotron DESY, Notkestr. 85, 22607 Hamburg, Germany}
\affiliation{Present address: European X-ray Free-Electron Laser, Holzkoppel 4, 22869 Schenefeld, Germany}
\author{Young Yong Kim}
\affiliation{Deutsche Elektronen-Synchrotron DESY, Notkestr. 85, 22607 Hamburg, Germany}
\affiliation{Present address: Pohang Accelerator Laboratory, Pohang 37673, Republic of Korea}
\author{Seon Woo Lim}
\affiliation{Department of Physics, Ulsan National Institute of Science and Technology, 50 UNIST-gil, Ulju-gun, Ulsan 44919, Republic of Korea}
\author{Daewoong Nam}
\affiliation{Pohang Accelerator Laboratory, Pohang 37673, Republic of Korea}
\affiliation{Photon Science Center, Pohang University of Science and Technology, Pohang 37673, Republic of Korea}
\author{Intae Eom}
\affiliation{Pohang Accelerator Laboratory, Pohang 37673, Republic of Korea}
\affiliation{Photon Science Center, Pohang University of Science and Technology, Pohang 37673, Republic of Korea}
\author{Minseok Kim}
\affiliation{Pohang Accelerator Laboratory, Pohang 37673, Republic of Korea}
\author{Sangsoo Kim}
\affiliation{Pohang Accelerator Laboratory, Pohang 37673, Republic of Korea}
\author{Ruslan Khubbutdinov}
\affiliation{Deutsche Elektronen-Synchrotron DESY, Notkestr. 85, 22607 Hamburg, Germany}
\author{Li Xiang}
\affiliation{Department of Chemistry, Pohang University of Science and Technology, Pohang 37673, Republic of Korea}
\author{Hoyeol Lee}
\affiliation{Department of Chemistry, Pohang University of Science and Technology, Pohang 37673, Republic of Korea}
\author{Moonhor Ree}
\affiliation{Pohang Accelerator Laboratory, Pohang 37673, Republic of Korea}
\affiliation{Department of Chemistry, Pohang University of Science and Technology, Pohang 37673, Republic of Korea}
\affiliation{Surface Technology Institute, Ceko Corporation, 519 Dunchon-daero, Jungwon-gu, Seongnam 13216, Republic of Korea}
\author{Chae Un Kim}
\affiliation{Department of Physics, Ulsan National Institute of Science and Technology, 50 UNIST-gil, Ulju-gun, Ulsan 44919, Republic of Korea}
\author{Ivan A. Vartanyants}
\email[corresponding author:\,]{ivan.vartaniants@desy.de}
\affiliation{Deutsche Elektronen-Synchrotron DESY, Notkestr. 85, 22607 Hamburg, Germany}
%

\date{\today}



\begin{abstract}

Understanding melting in metals is a hot topic of present research.
This may be accomplished by pumping the system with infrared (IR) laser radiation, and probing it with hard X-rays produced by an X-ray Free-Electron Laser (XFEL).
In this work we studied nanostructured polycrystalline thin films of platinum that were illuminated by IR radiation of increasing fluences.
We characterized the structural response as well as the nucleation and propagation of the liquid phase as a function of time delay between the IR pump and X-ray probe.
We observed partial melting of the samples for IR fluences higher than 200 mJ$\cdot$cm$^{-2}$.
To fit the contribution of the liquid phase to the scattering pattern in platinum we applied a model of liquid metal.
The two-temperature model simulations were performed to understand the solid-state fraction of the sample heating process as a function of time delay and fluence.

%
\end{abstract}


\maketitle

%
%
\section*{\label{sec:Introduction}Introduction}
When infrared (IR) light from a femtosecond laser interacts with free electrons in metals, it rapidly raises their kinetic energy, which is then partially transferred to the atomic lattice via electron-phonon coupling \cite{Anisimov1997}.
This excess of energy perturbs the atomic arrangement, whose temporal evolution can be probed, for example, by hard X-ray pulses generated by  ultrashort X-ray Free-Electron Lasers (XFELs) \cite{Emma2010,Ishikawa2012,Kang2017,Decking2020,Prat2020}.
Due to the exceptional properties in terms of brightness, temporal pulse duration and transverse coherence \cite{Vartanyants2011,Khubbutdinov2021,Kim2022}, hard XFELs are in fact unique tools to characterize structure and dynamics of condensed matter \cite{Clark2013,Bergeard2015,Schiwietz2016,Mukharamova2020}.

One important metal for pump-probe experiments, performed, for example, at XFELs, is platinum.
Platinum is a noble metal with numerous scientific and technological applications.
It is used, for example, for medical devices \cite{Cowley2011}, in nanomedicine \cite{Pedone2017} and has a central role in cancer chemotherapy \cite{Wang2005, Kelland2007}.
It is essential for the catalysis of the hydrogen oxidation and oxygen reduction reactions, at the core of fuel cells and electrolysers  \cite{Wu2013,Sheng2015,Fan2021}.
Due to its excellent performance and thermal stability, platinum is also extensively employed within automotive catalytic converters \cite{Sharma2012,YoungYongKim2021}, and to catalyse several other industrially-relevant reactions, for example, in the oil industry \cite{Hughes2021}.
The elevated chemical inertness, oxidation resistance, melting temperature and phase stability of platinum \cite{Anzellini2019} make it perfect to produce melting equipment.
For example, it is used to manufacture glass fibres, for temperature sensors or for space industry applications \cite{Fischer2001}.
Because of its properties, platinum is also used as reference for several applications, ranging from electrodes and thermometers to pressure calibrants \cite{Fratanduono2021}.

Pump-probe investigations 
have been performed earlier on aluminum  \cite{Williamson1984, Siwick2003} and gold (for example, evidence for bond hardening \cite{Ernstorfer2009}, identification of heterogeneous to homogeneous melting transition \cite{Mo2018} and elucidation of the role of grain boundaries on melting \cite{Assefa2020}).
Gold is the most widely studied element and possesses an atomic electronic configuration similar to the one of platinum, which differs from the former by only one electron.
However, theoretical calculations predict for platinum a completely different behavior in terms of electron-phonon coupling, and in turn of electron-lattice energy exchange \cite{Lin2008}.

With our experiment and its interpretation, we aim at increasing the understanding of platinum properties, and in particular of the crystalline-liquid phase transition in nanostructured thin films.
Furthermore, given platinum phase stability and resistance to oxidation, we aspire at using this experiment as a model for the melting process, whose detailed understanding is of great scientific importance.
Apart from increasing the comprehension of platinum and addressing thus a fundamental problem in material science, that is the nucleation and propagation of a liquid phase in films characterized by nanostructured domains, the presented results are potentially of technological relevance, given the increasing demand of ultrafast laser manufacturing \cite{Malinauskas2016,Vorobyev2013,Vorobyev2009,Green2014,Li2021,Semaltianos2010}.
%


%
\section*{\label{sec:Results}Results}
Platinum films of 100 nm thickness were grown on silicon nitride membranes by electron beam evaporation.
The resulting lateral dimension of the grains was of the order of 10 nm and they were preferentially columnar with respect to the film surface 
.

\begin{figure*}[ht]
\centering
\includegraphics[width=0.8\textwidth]{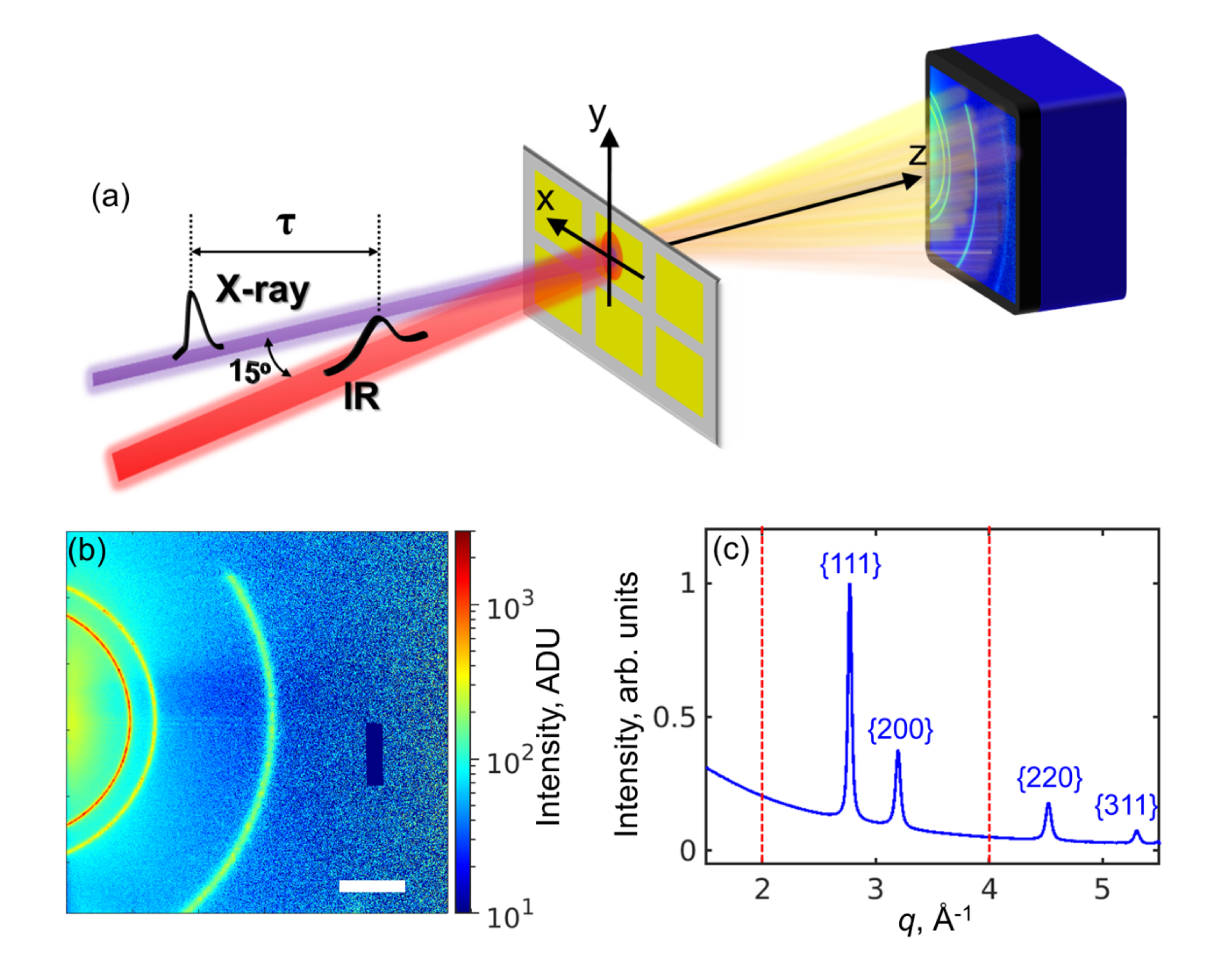}
\caption{%
\label{Figures:Figure_1}
(a) Schematic of the experimental setup.
The IR laser impinges on the sample 15 degrees apart from the X-ray beam (that is perpendicular to the sample) on the horizontal plane.
An on-line spectrometer, consisting of a curved Si crystal in (333) scattering geometry, provides a spectrum of each XFEL 
The silicon frame is moved in the xy-plane before each shot to illuminate a different window.
The X-ray scattering signal is measured on a pixelized detector positioned 108 mm downstream from the interaction region.
(b) Typical X-ray single-pulse scattering pattern of a platinum reference sample (that is the one without IR pump laser).
The scale bar corresponds to 500 pixels in detector units.
(c) Azimuthal average of the pattern in (b) as a function of momentum transfer where
Bragg peaks from the platinum fcc \{111\}, \{200\}, \{220\} and \{311\} family of planes can be observed.
Vertical dashed lines indicate the region that is used for further analysis of the pump-probe experimental data.
}%
\end{figure*}

The experiment was performed at the Nano-crystallography and Coherent Imaging instrument (NCI) of the Pohang Accelerator Laboratory X-ray Free-Electron Laser (PAL-XFEL) \cite{Park2016,Kang2017}.
A schematic of the experiment setup is shown in Fig. \ref{Figures:Figure_1}a.
The details of both the experiment and data analysis are provided in the Methods section.
%
%
%
%

\begin{figure*}[ht]
\centering
\includegraphics[height=0.75\textheight]{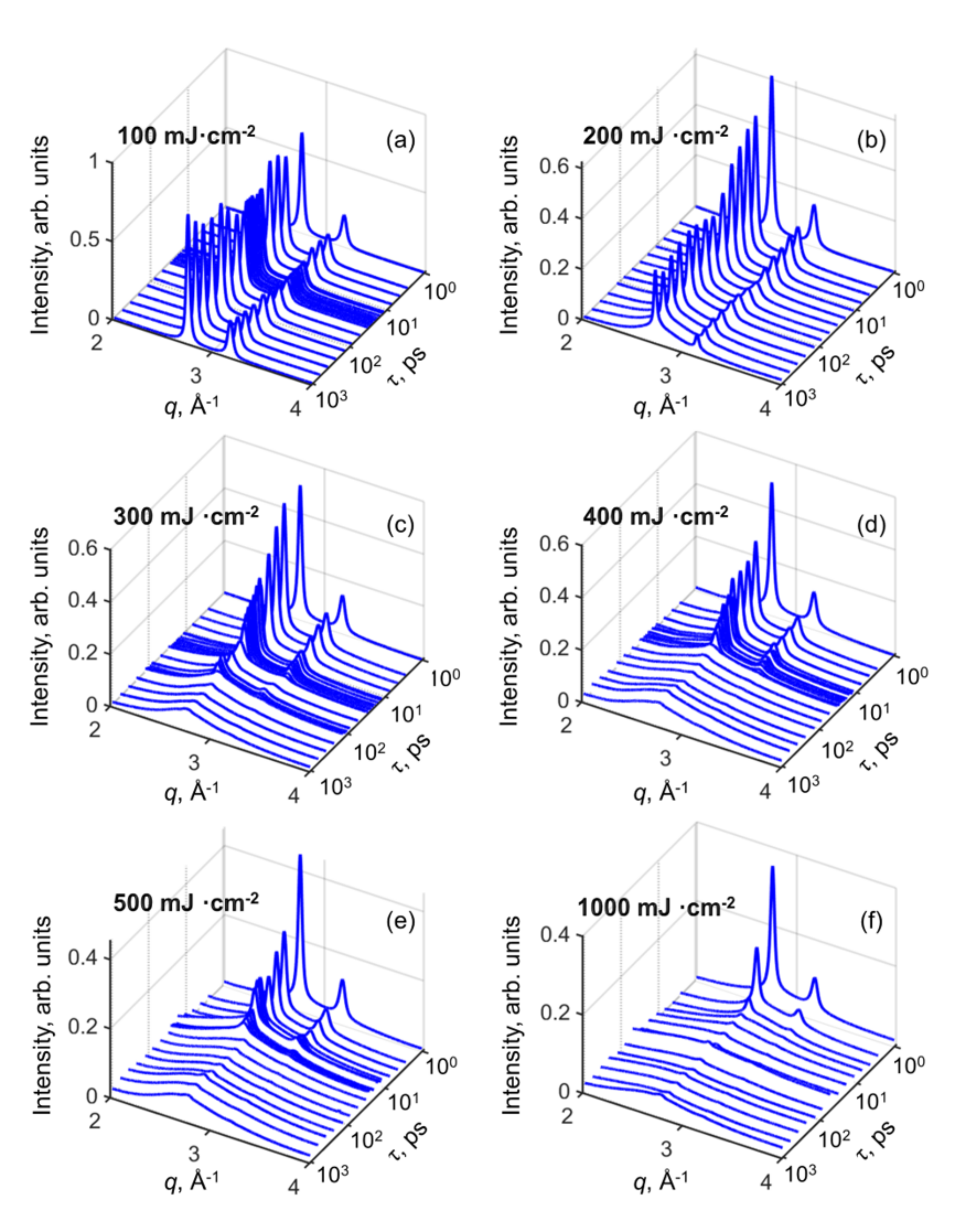}
\caption{%
\label{Figures:Figure_2}
Scattering curves at different time delays of the X-ray probe with respect to the IR pump for each IR fluence reported in Table \ref{tab:fluence}.
The data sets are normalized by the average maximum intensity of the reference sample, collected in absence of IR radiation.
}%
\end{figure*}

\begin{table}
\caption{%
\label{tab:fluence}Flux densities ($\Phi$) and corresponding pulse energies ($\mathcal{E}$, measured using a laser power meter) used to transfer energy to the sample. The estimated energy densities absorbed by the sample $\varepsilon$ are also reported.
}%
\begin{tabular}{|c|c|c|} \hline
   $\Phi$, mJ$\cdot$cm$^{-2}$ & $\mathcal{E}$, {\textmu}J & $\varepsilon$, kJ/kg \\ \hline\hline
   100 &  28.4 & 139 \\ \hline
   200 &  57.0 & 279 \\ \hline
   300 &  85.0 & 416 \\ \hline
   400 & 113.0 & 553 \\ \hline
   500 & 141.0 & 691 \\ \hline
 1,000 & 283.0 & 1,386 \\ \hline
\end{tabular}
\end{table}
The scattering curves collected during this experiment, with the subtracted background, are shown in Fig. \ref{Figures:Figure_2} for the different IR fluences used in this experiment.
%
%
%
By visual inspection, at the lowest IR fluence of 100 mJ$\cdot$cm$^{-2}$ no significant differences in the scattering data are observed at different time delays.
However, starting with the IR fluence of 200 mJ$\cdot$cm$^{-2}$ we observe a gradual decay of Bragg peak intensities with the increase of time delay.
While at this IR fluence both the (111) and (200) Bragg peaks corresponding to face-centered-cubic (fcc) Pt structure are visible for all time delays, already at 300 mJ$\cdot$cm$^{-2}$ the higher order peak (200) starts to disappear at about 100 ps.
In general, the decrease of intensities is faster the higher the IR fluence.
For example, at the fluence of 1,000 mJ$\cdot$cm$^{-2}$ Bragg peaks are practically vanishing after a few picoseconds from the IR pulse.
At the same time, we would like to note here that some intensity corresponding to the (111) Bragg peak is observed for every fluence and time delay, indicating that some residual solid fraction is always present in the region of the sample probed by X-rays.
In fact, we interpret the smooth continuous scattering curve which appears as the intensity of Bragg peaks decreases, as the fingerprint of the liquid platinum phase.
By that, we conclude that platinum films are partially melted for fluences higher than 200 mJ$\cdot$cm$^{-2}$.
%
Below, we will present a detailed analysis of the results obtained.

\begin{figure*}[ht]
\centering
\includegraphics[height=0.65\textheight]{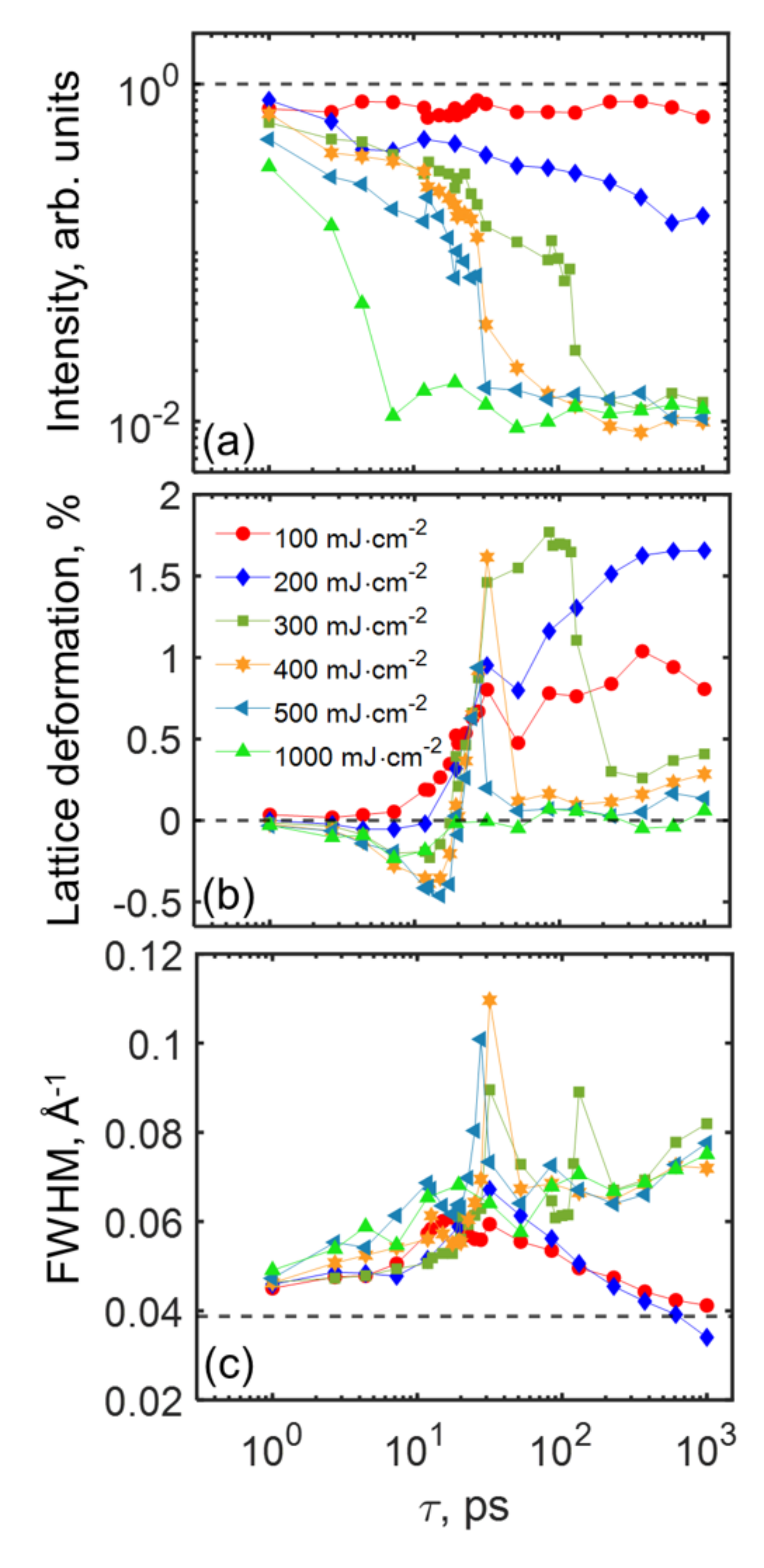}
\caption{%
\label{Figures:Figure_3}
Parameters deduced from the fitting of the (111) Bragg peak.
(a) Integrated intensity normalized to the reference value,
(b) lattice deformation (see text for details), and
(c) Full-Width at Half-Maximum (FWHM) of the (111) Bragg peak.
In all panels dashed lines are showing the values of parameters for the reference value in absence of IR radiation.
Data are plotted as a function of the time delay between the IR pump and X-ray probe.
}%
\end{figure*}

The parameters extracted from the fit of scattering curves are reported in Fig. \ref{Figures:Figure_3} as a function of pump-probe time delays.
Each point shown in this figure is the median of the corresponding dataset 
.
Even if both the (111) and (200) Bragg peaks are present in the investigated momentum-transfer range, we restrict our analysis below to the former, since the behaviour of the two is highly correlated within the resolution of our measurements 
.


We begin our discussion from the integrated intensity of the (111) Bragg peak (Fig. \ref{Figures:Figure_3}a), which is proportional to the amount of crystalline fcc platinum in the probed region of the sample.
As we can see from this figure, for each value of the incident IR fluence this quantity is reduced with respect to the reference value (that is measured only with XFEL radiation without illuminating the sample with IR).
As discussed above, the decrease within the investigated temporal range is negligible when the IR fluence is equal to 100 mJ$\cdot$cm$^{-2}$ and modest when it is 200 mJ$\cdot$cm$^{-2}$ (these will be named as "low fluences").
On the contrary, for higher values ("high fluences"), in our experiment, actually starting from 300 mJ$\cdot$cm$^{-2}$, the 
integrated intensity drops to about 1\% of the reference value, and the decay is faster with higher fluence.
This occurs from about one hundred picoseconds at 300 mJ$\cdot$cm$^{-2}$ to a few picoseconds for the 1,000 mJ$\cdot$cm$^{-2}$ case.

The position of a Bragg peak $q_{hkl}$ is related to the lattice parameter $a$ of a cubic lattice through the known relation $a = 2\pi \sqrt{h^2 + k^2 + l^2} / q_{hkl}$.
Variations of the strain value as a function of the time delay, defined as $\varepsilon(\tau)=\left[a(\tau)-a_0\right]/a_0$, are shown in Fig. \ref{Figures:Figure_3}b.
For low fluences, it can be observed that the lattice is expanded for each time delay longer than approximately 10 ps.
This is compatible with an increase of the sample temperature.
Assuming the expansion to be purely thermal, a maximum temperature increase at one nanosecond, at the end of our observation, was estimated to be about 1,300 K and 1,800 K for IR fluences of 100 mJ$\cdot$cm$^{-2}$ and 200 mJ$\cdot$cm$^{-2}$, respectively.
The thermal expansion coefficient was obtained from Ref.\cite{Kirby1991}.
The general trend for each IR fluence greater than 100 mJ$\cdot$cm$^{-2}$ is that the lattice is progressively becoming more compressed in the first 10 - 20 ps, up to approximately $\varepsilon=$-0.5\% at the IR fluence of 500 mJ$\cdot$cm$^{-2}$.
For a purely elastic deformation, with the Young's modulus equal to\cite{Cardarelli2008} 168 GPa this value corresponds to the stress value of about 0.8 GPa.
Compression is increasing with IR fluence, except for the 1,000 mJ$\cdot$cm$^{-2}$ case, likely due to the extremely small residual solid state fraction (see Fig. \ref{Figures:Figure_3}a).
After this initial phase, for each IR fluence the lattice starts to rapidly expand up to time delays of approximately 30 ps.
This is about the time necessary for a sound wave to transverse the thickness of the platinum sample, which is 100 nm in our case.
Indeed, if a deformation wave is traveling through the platinum film at the (longitudinal) speed of sound \cite{dePodesta2002} of 3,260 m/s, this time is 31 ps.
After this point, for low fluences the strain values decrease slightly up to time delays of about ~50 ps, and then they increase again.
For high fluences, in general, the expansion diminishes, following the same trend already observed for the residual fcc solid state fraction.

The 
FWHM of the 111 reflection is shown in Fig. \ref{Figures:Figure_3}c.
It is inversely proportional to the size of the coherently scattering domains and proportional to the microstrain and thermal vibrations in the sample \cite{Warren}.
For each value of the IR fluence, the peak width is initially slightly increased with respect to the reference value.
However, at about 30 ps we observe a steep increase of the peak width followed by a fast decay.
The curve has a maximum at $31.6$ ps for fluences lower than 500 mJ$\cdot$cm$^{-2}$.
This is likely due to extremely small residual fcc fraction at this time delay (see Fig. \ref{Figures:Figure_3}a), which might also explain why the peak width for the IR fluence of 1,000 mJ$\cdot$cm$^{-2}$ does not follow a similar trend.
It is interesting to note, that for low fluences, the FWHM at $1$ ns is practically returning to its initial value.
In particular, for the 200 mJ$\cdot$cm$^{-2}$ case the final FWHM value in the sub-nanosecond time delay region is even lower than the reference one, possibly indicating an increase of the crystallite size due to sintering of smaller grains or relaxation of residual stresses.
At the same time, for high fluences, the values of FWHM after the peak at 31.6 ps continue to increase.

%
%

The Kirchhoff's law predicts, for a constant pressure process,
the change of the temperature $T$ of the system as a function of enthalpy $\mathcal{H}$ to be $\mathrm{d}T = \mathrm{d}\mathcal{H} / c_p (T)$.
We assume also here that all the energy absorbed by the system is increasing its temperature (see Table \ref{tab:fluence}).
The above equation gives us an estimate of the sample temperature, it should reach a maximum temperature of ${\sim}1,300$ K, and therefore remain solid, when irradiated with an IR fluence of 100 mJ$\cdot$cm$^{-2}$
(we remind that the melting temperature for platinum at atmospheric pressure is $T_M$=2,041.3 K \cite{Bergeard2015}).
It should be noted that this is the same temperature estimated from the position of the (111) Bragg peak assuming ambient pressure.
For the IR fluence of 200 mJ$\cdot$cm$^{-2}$ the melting should be partial, with coexistence of the solid and liquid phases, whereas 300 mJ$\cdot$cm$^{-2}$ is sufficient to transform the solid into liquid.
This picture is apparently consistent with Fig. \ref{Figures:Figure_3}a for long time delays although an extremely small residual fcc solid state fraction is always present according to our experimental findings.
We speculate that this might be due to chemical stabilization of platinum at its interface with Si$_3$N$_4$ film, inhibiting crystal to liquid phase transition.

%
%

The scattering curve attributed to liquid, which is clearly distinguishable from the Bragg peaks, was fitted using the model developed by Ashcroft \cite{Ashcroft1966, Pedersen1994} (see Methods section).
It should be noted that, especially for low fluences and short pump-probe delays, this model is likely to fit the diffuse scattering due to various defects including thermal vibrations, and might therefore not provide correct information on the liquid phase.
The area under the broad Ashcroft's curve, which is proportional to the amount of liquid phase, was estimated by integrating the corresponding analytical model from the origin of momentum-transfer space to infinity.
To compensate for fluctuations of the incoming X-ray intensities, we divided the integral by the total intensity in the background for each diffraction pattern.

\begin{figure}[ht]
\centering
\includegraphics[height=0.65\textheight]{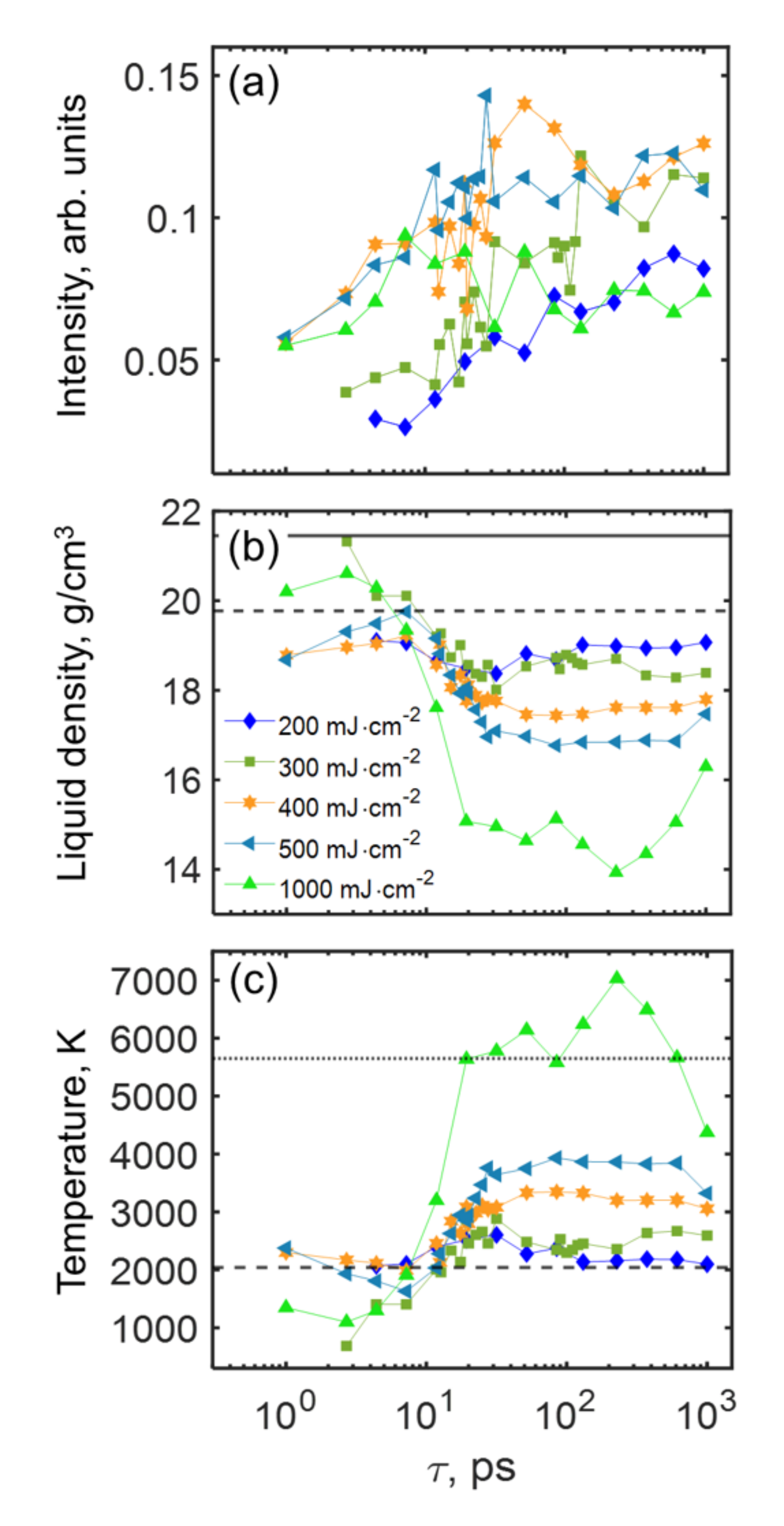}
\caption{%
\label{Figures:Figure_4}
Quantities extracted from the liquid phase model (see text for details).
(a) Integrated liquid phase intensity normalized to the background intensity,
(b) liquid density, and
(c) temperature.
In (b) the dashed and solid lines indicate the nominal values for the liquid and solid density, respectively and in (c) dashed and dotted lines indicate the melting and boiling temperatures, respectively.
Data are plotted as a function of the time delay between the IR pump and X-ray probe.
We excluded the lowest fluence of 100 mJ$\cdot$cm$^{-2}$ at which no liquid phase was detected.
}%
\end{figure}

The result is shown in Fig. \ref{Figures:Figure_4}a.
The general trend is that the integral of this liquid
curve increases as a function of the time delay, indicating an increase of the liquid fraction.
It also increases as a function of the IR fluence, although there is no significant difference between 400 mJ$\cdot$cm$^{-2}$ and 500 mJ$\cdot$cm$^{-2}$.
Moreover, at time delays of hundreds of picoseconds the values of integrated intensities for the IR fluence of 1,000 mJ$\cdot$cm$^{-2}$ are similar to the ones for 200 mJ$\cdot$cm$^{-2}$.
%
%
To explain this behaviour, we evaluated the conditions for ablation using the approach described in Ref.\cite{Gamaly2002} 
.
According to this, the estimated crater depth is about 1 nm and 10 nm for the IR fluences of 500 mJ$\cdot$cm$^{-2}$ and 1,000 mJ$\cdot$cm$^{-2}$, respectively 
.

In the model developed by Ashcroft \cite{Ashcroft1966, Pedersen1994}, the profile of the scattering curve of a liquid of hard spheres is defined by their diameter $d$ and packing density $\eta$.
The two parameters are in turn related to the liquid density of platinum $\rho_{Pt}$ by the relation $\rho_{Pt} = 6 \pi \eta A_{Pt} / d^3$, where $A_{Pt} = 195.08$ g/mol is the atomic weight of platinum.
%
The density of the liquid phase, shown in Fig. \ref{Figures:Figure_4}b, does not show an obvious trend for time delays shorter than about 10 ps.
As we can see from the same figure, in the same region of time delays, the liquid density is lower the higher the IR fluence. This observation is compatible with an increase of the liquid temperature.
However, if we exclude the data for 200 mJ$\cdot$cm$^{-2}$ and 300 mJ$\cdot$cm$^{-2}$, associated with a small liquid fraction (see Fig. \ref{Figures:Figure_4}a), then the higher the fluence the less dense is the liquid.
We speculate that this might be due to the residual solid fraction (see Fig. \ref{Figures:Figure_3}a) mechanically constraining the liquid phase.
On the contrary, for all IR fluences and time delays larger than approximately 10 ps, the liquid density is lower than the reference value at the melting point\cite{Mehmood2012} of 19.77 $g/cm^3$.
In the same region of time delays, it is also lower the higher is the IR fluence, compatibly with an increase of the liquid temperature.

The temperature of the liquid phase can be estimated from Ref.\cite{Hixson1993} as $T = (V/V_0 - 0.89125)/8.3955 \times 10^{-5}$, where the volume $V$ is calculated from the density as obtained from the Ashcroft model, and $V_0 = 4.673 \times 10^{-5}$ $m^3 \cdot kg^{-1}$.
It should be noted that this equation is valid in the interval of temperatures from 2,042 K to 5,650 K, and it assumes a constant pressure process.
Values of temperature of the liquid phase, shown in Fig. \ref{Figures:Figure_4}c, outside this range are in general the ones corresponding to time delays shorter than 10 ps, or to the highest IR fluence of 1,000 mJ$\cdot$cm$^{-2}$.
The values within the range of validity of the equation, in Fig. \ref{Figures:Figure_4}c, confirm that the temperature of the liquid phase increases with the IR fluence.
In particular, it is approximately the melting temperature for the IR fluence of 200 mJ$\cdot$cm$^{-2}$, the temperature at which solid and liquid coexist, and it reaches almost 4,000 K for the 500 mJ$\cdot$cm$^{-2}$ case.
Interestingly, the temporal region between approximately 10 and 30 ps corresponds to both a transition between two different states of density and temperature, and of strain in the lattice (see, for example, Fig. \ref{Figures:Figure_3}b), suggesting that the solid and liquid phase might experience the same mechanical pressure.

To support the interpretation of experimental results, we performed two-temperature model (TTM) simulations\cite{Anisimov1974} 
.
The TTM is a phenomenological description of the spatio-temporal evolution of the temperature in a metal following its ultrafast excitation.
In this picture, the laser pulse is absorbed by free electrons in a thin layer of about 10 nm and rapidly increases their energy.
This is then partially transferred to the lattice via electron-phonon coupling.
The TTM assumes a continuous media and does not include any defects, like the grain boundaries, present in the investigated samples.
These grain boundaries can act as nucleation seeds for the liquid phase, lowering the amount of energy required for the phase transition (heterogeneous melting)\cite{Assefa2020}.

\begin{figure*}[ht]
\centering
\includegraphics[height=0.55\textheight]{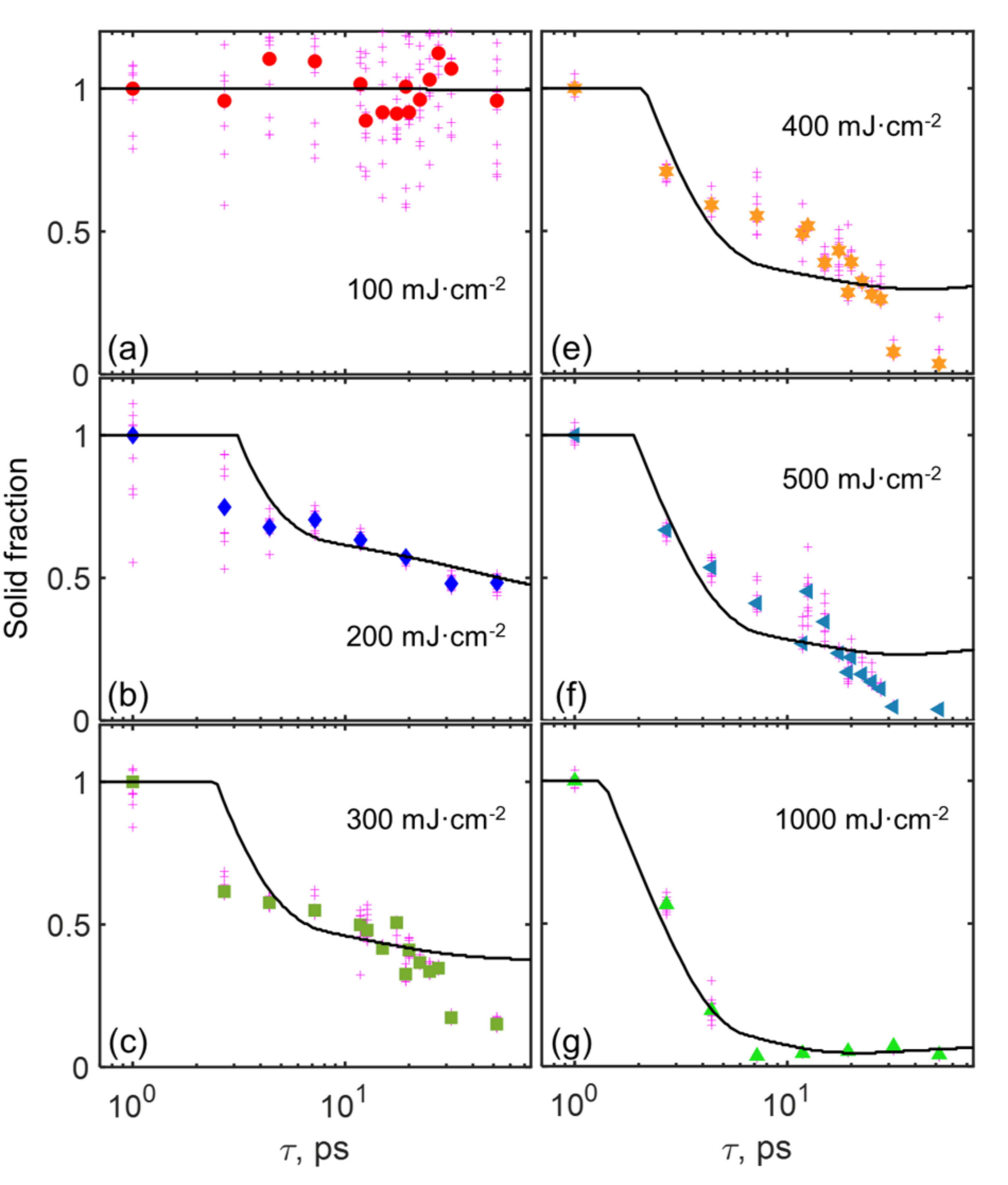}
\caption{%
\label{Figures:Figure_5}
Solid state fraction.
Dots are the median values for a given time delay and pink crosses are the values calculated from the experimental data.
The solid black line is calculated from the TTM simulations.
}%
\end{figure*}

In Fig. \ref{Figures:Figure_5} we compare the fraction of fcc solid state estimated from the experimental data and the one obtained from the TTM model.
The values of experimental data are calculated assuming that the integral of the (111) Bragg peak is proportional to the amount of solid phase, and that there is no liquid platinum at the time delay of 1 ps.
The value of the peak integral at this time delay is, therefore, used to normalize all the other values.
For the case of the TTM simulations, the fraction of solid is simply the thickness of the solid phase at a given time from the IR pulse divided by the total thickness of the simulation domain (100 nm).
It is important to note here that we did not introduce any fitting parameters in the TTM simulations.
All parameters, including the electron-phonon coupling parameter $G_{ep}$, were taken from the known references 
.
As can be observed from Fig. \ref{Figures:Figure_5}, the model qualitatively captures the trends observed experimentally.
For the IR fluence of 100 mJ$\cdot$cm$^{-2}$ it does not predict any formation of the liquid phase, which is compatible to our experimental findings.
From the fluence of 200 mJ$\cdot$cm$^{-2}$ the liquid phase is formed starting from about a few picoseconds.
The time delay for the formation of the liquid phase is becoming shorter with increasing the IR fluence.
For example, for the 1,000 mJ$\cdot$cm$^{-2}$ case the solid fraction is transformed to liquid already at time delays of about 2 ps.
For all fluences, experimental results seem to indicate a faster nucleation of the liquid phase with respect to simulations. This could either be because our TTM model does not include any structural defects, which are present in the samples (homogeneous versus heterogeneous melting), or because some of the simulation parameters are not accurate.
In particular, a relevant role in defining the temporal evolution of the atomic lattice is played by the electron-phonon coupling.
Interestingly, our simulations show, that increasing/reducing the electron-phonon coupling parameter by twice is reducing/increasing the start of the melt process by 10 ps.
For IR fluences from 300 mJ$\cdot$cm$^{-2}$  to 500 mJ$\cdot$cm$^{-2}$ we observe a second drop of the experimental solid fraction at about 20 - 30 ps, that is not explained by the TTM.
This drop might be related to propagation of the sound wave through the 100 nm Pt sample layer.
At the same time, we observe excellent agreement between the solid fraction calculated from experimental data and the TTM for the highest IR fluence of 1,000 mJ$\cdot$cm$^{-2}$ , although, we recall, the sample is partially ablated.
In summary, we described the evolution of platinum nanostructured films illuminated by ultrashort IR radiation at several fluences and time delays.
We observed a deformation wave traveling perpendicular to the film at the speed of sound, and the temperature of the films increasing as a function of time delay.
The sample remained solid when the fluence was 100 mJ$\cdot$cm$^{-2}$, and it was partially liquid for higher fluences.
We observed that lattice deformations before the sound wave travelled through the entire sample were mostly compressive and after that they were causing expansion.
Analysis of our results with the Ashcroft's model suggested that the density of the liquid phase was dropping after the sound wave propagated through the sample.

To support experimental findings, we performed TTM simulations.
They predicted the sample to be solid at the IR fluence of 100 mJ$\cdot$cm$^{-2}$, and partially liquid for higher fluences, with the solid fraction decreasing the more, the higher the fluence.
We also observed an additional drop of the solid fraction that is not described by TTM simulations at fluences from 300 mJ$\cdot$cm$^{-2}$
to 500 mJ$\cdot$cm$^{-2}$ and time delays longer than 30 ps.
This is most probably related to propagation of the sound wave through the 100 nm platinum sample layer.
Our TTM simulations, based on a model of a uniform layer, are not sufficient to capture the complexity of the solid-liquid phase transition in polycrystalline films illuminated by femtosecond IR light.
In particular, the ablation process or the presence of defects such as grain boundaries are not taken into account in TTM model.

Finally, we believe that a deep understanding of the results presented here will need more extensive experimental
and theoretical exploration in the future.
%
We think that research performed at XFEL facilities will be instrumental in answering questions outlined in this work.

\section*{\label{sec:Methods}Methods}
\subsection*{\label{subsec:Experiment}Experiment}
%
The X-ray beam propagating perpendicular to the sample surface was focused using a Kirkpatrick-Baez mirror system \cite{Kirkpatrick1948} on an area of approximately 5 $\mu$m in diameter on the sample 
.
The average X-ray photon energy during the experiment, measured by an on-line spectrometer, was 9.7 keV 
.
The resolution of the on-line spectrometer was estimated to be $0.26$ eV Full-Width at Half-Maximum (FWHM) \cite{Nam2021}.
Scattering data were collected on a Rayonix MX225-HS detector (2,880${\times}$2,880 pixels of size 78${\times}$78 {\textmu}m$^2$ with a 2${\times}$2 binning mode).
The detector was positioned downstream from the interaction region at a distance of $108$ mm and its center was shifted from the X-ray path by $-37$ mm and $1$ mm along the $x-$ and $y-$axes, respectively (the reference frame is shown in Fig. \ref{Figures:Figure_1}a).
The sample-detector distance was determined by fitting platinum Bragg peaks of the reference sample and assuming the lattice constant to be 3.9236 {\AA} \cite{Arblaster1997}.
A typical scattering pattern measured by the detector is shown in Fig. \ref{Figures:Figure_1}b.
Before each shot, the sample was moved in the plane perpendicular to the X-ray beam (the $xy$-plane in Fig. 1a) so to illuminate a different window
.
%
%

A Ti:sapphire laser (800 nm central wavelength) \cite{Kim2019}, generating 100 fs (FWHM) pulses and focused on an area of about 110 {\textmu}m (FWHM) 
, was used to increase the temperature of the samples.
The direction of the IR pulses was 15$^\circ$ from the sample normal on the horizontal plane (see Fig. \ref{Figures:Figure_1}a).
The values of IR laser fluences, reported in Table \ref{tab:fluence}, were varied from 100 mJ$\cdot$cm$^{-2}$ to 1,000 mJ$\cdot$cm$^{-2}$.
The time-delay between the IR pump and X-ray probe was varied from 1 ps to 1,000 ps with logarithmic sampling.
At least 10 patterns were collected for each combination of fluence and time-delay, moving to a different window after each pulse.
Before each X-ray pulse, the sample position along the X-ray beam ($z$-axis) was adjusted in order to optimize the focal plane as seen by an inline microscope.
Given the uncertainty on the focal plane position and a drift of time zero across the experiment of about 0.5 ps, the error in time delay was estimated to be of the order of 1 ps.

%
\subsection*{\label{sec:Data Analysis}Data analysis}

The first step of the analysis of scattering data was the removal of the pedestal (dark signal) and bad pixels.
Each frame was further corrected by normalizing data by solid angle, X-ray polarization, linear absorption of X-rays by the sample, and air scattering in the sample-detector path 
.
%
Some regions of the detector were masked because the scattered intensity was 
shadowed by the sample holder (see, for example, Fig. \ref{Figures:Figure_1}b 
).
A few detector frames (less than 7\% of the total) were considered as outliers and discarded, for example, when the X-ray beam was too close to an edge of the silicon frame.
The average and standard deviation of the radial profile of each image were finally computed using the wavelength deduced from the in-line spectrometer.
The momentum transfer axis ($q$) was binned with a bin size of 0.005 {\AA}$^{-1}$.
A typical radial profile for the reference sample, obtained by applying the above-mentioned pipeline to the detector images in Fig. \ref{Figures:Figure_1}b, is shown in Fig. \ref{Figures:Figure_1}c.
High-$q$ data were typically noisy and therefore the momentum range for further analysis was restricted to the interval from 2 \AA$^{-1}$ to 4 \AA$^{-1}$ (see Fig. \ref{Figures:Figure_1}c).
In this region, face-centered cubic (fcc) platinum has two Bragg reflections, the one associated to the $\{111\}$ ($q-$value of 2.774 {\AA}$^{-1}$) and $\{200\}$ ($q-$value of 3.203 {\AA}$^{-1}$) family of planes.
To account for the contribution of scattering of X-rays by air in the optical path and the silicon nitride substrate, a membrane (no sample) was measured and fitted with a 4-th order Chebyshev polynomial of the first kind.
The result of the fit, $\mathcal{B}_0$, was then used to construct a general background function $\mathcal{B}(q) = \alpha \mathcal{B}_0 \mathrm{exp} \left(\kappa q\right)$.

To fit the scattering curves we defined the following model as the sum of three contributions: (i) the background $\mathcal{B}(q)$ (2 free parameters: $\alpha$ and $\kappa$), (ii) a pseudo-Voigt function for each Bragg peak (4 free parameters: the central position $q_{(hkl)}$, FWHM, area of the curve, and the Lorentzian fraction) and (iii) the melting model proposed by Ashcroft \cite{Ashcroft1966, Pedersen1994} for the liquid signal (3 free parameters).
Ashcroft derived the liquid structure factor of a system of hard-spheres as a function of their diameter and packing density \cite{Ashcroft1966}.
We additionally multiplied the structure factor by a scaling parameter and the squared modulus of the platinum atomic form factor (coefficients were obtained from Ref. \cite{Brown2006}).
%

\vspace{1\baselineskip}

\textbf{Competing Interests}

The authors declare that they have no competing interests.

\vspace{1\baselineskip}

\textbf{Funding}

LG, YYK, RK, and IAV acknowledge the support by the Helmholtz Associations Initiative and Networking Fund.
DN was supported by the NRF of Korea (Grant No. 2021R1F1A1051444 and 2022M3H4A1A04074153).
SWL and CUK acknowledge the support by the National Research Foundation of Korea (NRF) grant funded by the Korea government (NRF-2022R1A2C2091815).

\begin{acknowledgments}
The use of the Maxwell computational resources operated at Deutsches Elektronen-Synchrotron (DESY) is acknowledged.
The authors would like to thank S. Lazarev for the stimulating discussions while planning the experiment.
The authors acknowledge fruitful discussions with M. Altarelli and I.K. Robinson.
The authors acknowledge careful reading of the manuscript by H.-P. Liermann and G. Hinsley.
\end{acknowledgments}

\bibliography{bibliography}

\eject

\end{document}